\documentclass[]{spie}  %>>> use for US letter paper
\def\code#1{\texttt{#1}}

\usepackage{amsmath,amsfonts,amssymb}
\usepackage{graphicx}
\usepackage[colorlinks=true, allcolors=blue]{hyperref}
\usepackage{todonotes}

\title{The Roman Coronagraph Community Participation Program: data reduction pipeline astrometric calibration}

\author[1,2]{Amanda Chavez}
\author[1,2]{Jason Wang}
\author[3]{Vanessa P. Bailey}
\author[4]{Maxwell Millar-Blanchaer}
\author[5]{Julien Girard}
\author[3]{John Krist}
\author[3]{Julia Milton}
\author[3]{Marie Ygouf}

\affil[1]{Northwestern University, 633 Clark Street Evanston, IL 60208, USA}
\affil[2]{Center for Interdisciplinary Exploration and Research in Astrophysics, 1800 Sherman Ave, Evanston, IL 60201, USA}
\affil[3]{Jet Propulsion Laboratory, California Institute of Technology, 4800 Oak Grove Drive, Pasadena, CA 91109, USA}
\affil[4]{University of California Santa Barbara, Santa Barbara, CA 93106, USA}
\affil[5]{Space Telescope Science Institute, 3700 San Martin Drive, Baltimore, MD 21218, USA}

\authorinfo{Further author information: (Send correspondence to Amanda Chavez)\\Amanda Chavez: E-mail: amchavez@u.northwestern.edu\\  Jason Wang: E-mail: jason.wang@northwestern.edu}

\begin{document} 
\maketitle

\begin{abstract}
% should be no more than 250 words 
The Nancy Grace Roman Space Telescope will be equipped with a Technology Demonstration Coronagraph Instrument that will push the current limits of high contrast imaging for exoplanets ($\sim10^{-9}$ contrast). The Roman Coronagraph Community Participation Program has developed \code{corgidrp}, a python-based data reduction pipeline for the Roman Coronagraph Instrument that will perform essential data processing and calibration steps for coronagraphic observations. The astrometric calibration function within \code{corgidrp} allows us to understand the on-sky angular size and distance scale of science observations by characterizing essential detector parameters: boresight, plate scale, north angle, and optical distortion. Measuring these astrometric calibration products not only helps us understand the science output of our data, but it is what allows us to point the Roman coronagraph accurately at our science target in the first place. Here, we describe the techniques used within the astrometric calibration and demonstrate that our algorithm meets Technology Demonstration Threshold Requirements: (1) compute the on-sky location of the center of CGI EXCAM detector to better than 30 [mas] and (2) compute the on-sky position angles of the camera axes to within 0.3 [deg]. 
\end{abstract}


\section{INTRODUCTION}\label{sec:intro}

The upcoming Roman Space Telescope is a 2.4 meter observatory named after NASA's first Chief of Astronomy, Dr. Nancy Grace Roman. The telescope will be equipped with a Wide Field Imager and a Technology Demonstration Coronograph Instrument (CGI). The coronagraph will push the current limits of high contrast direct imaging with the goal of reaching $\sim10^{-9}$ contrast ratios at visible and near-IR wavelengths from 575[nm] to 825[nm]\cite{Kasdin_2020, Bailey_2023}. The Roman CGI will test high contrast imaging technology in preparation for the next generation of space telescopes, such as Habitable Worlds Observatory, that aim to image Earth-like planets around Sun-like stars (contrast ratios $\sim10^{-10}$) in visible wavelengths\cite{decadal_survey_2020}. 

The Roman Community Participation Program (CPP) has developed the Coronagraph Instrument Data Reduction Pipeline\cite{Millar_Blanchaer_Wang_2024, Wang_2026} (\code{corgidrp}), to process CGI data. \code{corgidrp} is a python-based processing pipeline that provides a streamlined procedure for translating raw coronagraphic data into decipherable science products. \code{corgidrp} is built to ensure that CGI data can meet all formal Technology Demonstration Threshold Requirements. In this proceedings, we turn our focus to the astrometric calibration function within this data reduction pipeline.

The \code{corgidrp} astrometric calibration function is crucial for both instrument operation and for scientific observations. This calibration process allows us to understand several key characteristics of the CGI detector, namely: the pointing offset, the orientation, the pixel scale, and the optical distortion. These data products are essential for instrument operations because they calibrate the commanded CGI pointing and allow us to successfully acquire science targets. This calibration must be done quickly and accurately in order for CGI to efficiently acquire a star and continue with commissioning activities. Ultimately, the astrometric calibration allows us to interpret the angular size and distances of sources within our science observations. Particularly, astrometric calibration plays a crucial role for obtaining the precise astrometry of reflected light planets, as their brightness depends heavily on their orientation with respect to the host star. For these reasons, we are motivated to produce an accurate and efficient astrometric calibration algorithm for the Roman Coronagraph Instrument. 

In this proceedings, we detail the key processes of the astrometric calibration and share the results of our testing. In \hyperref[sec:finding_and_matching]{Section 2}, we share our algorithm for detecting sources in a CGI detector image and matching them to real sky coordinates. In \hyperref[sec:astrometric_calibration]{Section 3}, we outline the processes for calculating the following astrometric calibration products: plate scale in \hyperref[subsec:platescale]{Section 3.1}, north angle in \hyperref[subsec:northangle]{Section 3.2}, boresight in \hyperref[subsec:boresight]{Section 3.3}, and the distortion map in \hyperref[subsec:distortion]{Section 3.4}. In \hyperref[sec:testing]{Section 4}, we describe our testing techniques and results.

\section{AUTOMATED SOURCE MATCHING}\label{sec:finding_and_matching} 

The astrometric calibration begins with a CGI detector image that has an unknown pointing offset (relative to commanded [RA, Dec] pointing), unknown magnification, and unknown orientation. The challenge of the astrometric calibration is to orient the detector image with the reference field it is observing. To do this, we rely heavily on accurately matching stars from their [pixel, pixel] position on the detector to their true [RA, Dec] coordinates. We choose to calibrate the Roman CGI to the James Webb Space Telescope's calibration field\cite{Zellem_2022} and, therefore, use the Large Magellanic Could (LMC) as our reference field. The LMC is a high density field of uniformly distributed bright stars and is large enough such that a significant offset in commanded CGI pointing will not steer us outside of the field-- proving it to be an ideal astrometric calibration reference. 

The automated source finding and matching algorithm within the astrometric calibration allows us to accurately map the LMC stars from their detector positions to their true sky coordinates. The results of this matching process provide a basis for computing all astrometric calibration products. The first step of our algorithm is to locate several ($\sim15$) bright sources in a CGI detector image. We filter the detector image for bright pixels using the \code{numpy}\cite{numpy_2020} maximum function, which returns the single brightest pixel across the entire image. We then utilize \code{pyklip}\cite{Wang_2015} to fit a 2D gaussian at the brightest pixel's [x, y] coordinates-- allowing us to obtain subpixel precision when recording the final source position. Then, we mask out the point spread function (PSF) of the bright source assuming a tunable pixel radius ($\sim7$ [pixels]). Finally, we repeat the process until N bright sources are found. 

\begin{figure}[ht]
   \begin{center}
   \begin{tabular}{c} 
   \includegraphics[height=8cm]{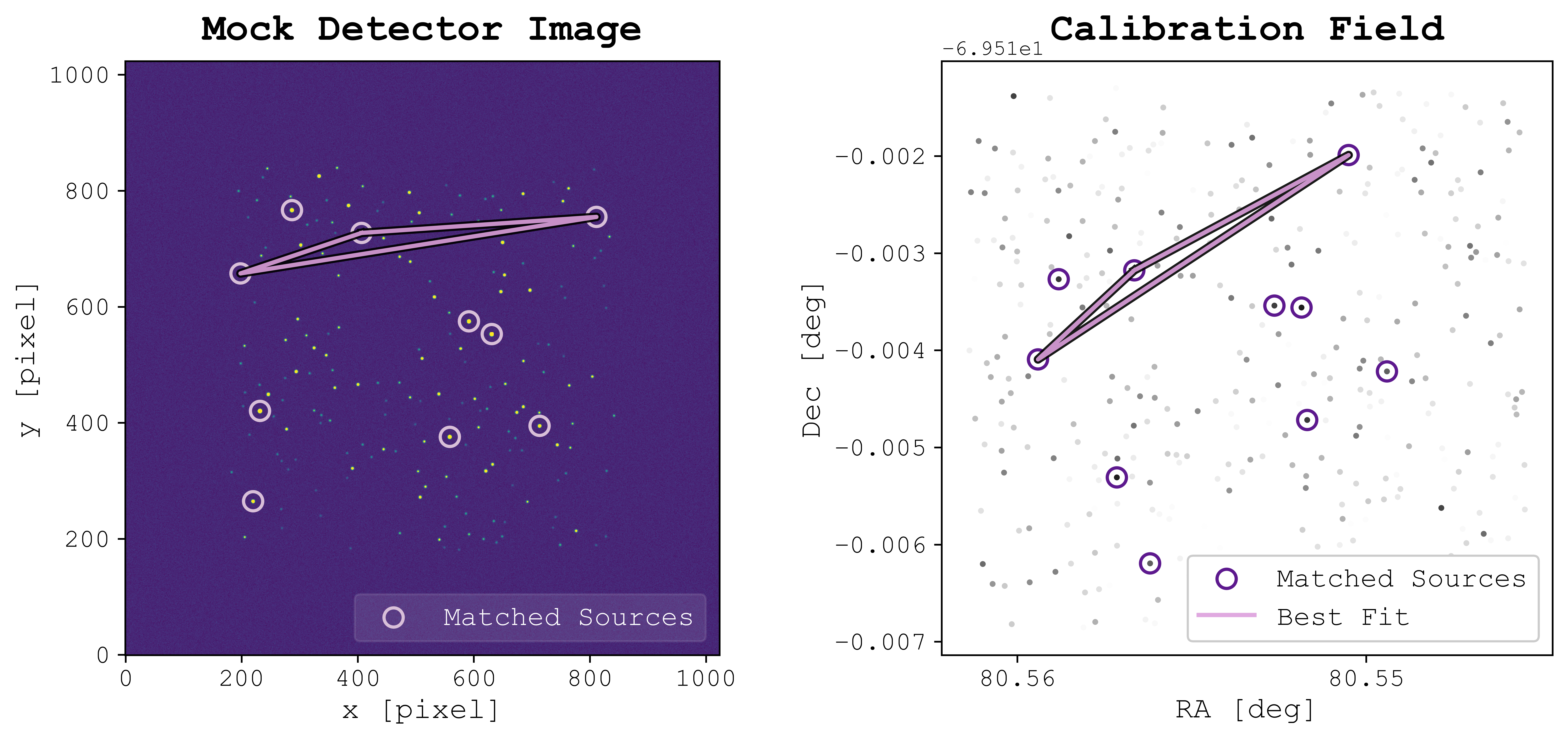}
   \end{tabular}
   \end{center}
   \caption{\label{fig:matching} 
\textbf{Left:} A mock CGI detector image including 10 sources found by our automated source matching algorithm (purple circles). The initial triangle of 3 bright sources used for matching is outlined. \textbf{Right:} The LMC reference subfield with the best fit source triangle outlined and all other matched sources circled.}
\end{figure}

The second step of the algorithm matches the CGI detector sources to their [RA, Dec] sky coordinates. We take the brightest three sources in the CGI detector image as a triangle and compute the ratio of each side length to the perimeter-- creating a unique characterization of the three sources' geometry that requires no assumptions about the detector plate scale or north angle. To avoid error in matching, we restrict our calibration field to a $\sim0.72$ [arcminute] radius centered at the detector's commanded pointing coordinates [RA, Dec]. This restriction allows us to consider sources more than twice the full detector FOV ($\sim7.5$ [arcsec] radius) away from the target pointing-- ensuring we do not miss the correct LMC star matches while also allowing us to reduce the number of stars we consider in the matching process. Then, we compute all possible combinations of 3 stars from N bright sources in this LMC calibration subfield (we default to using the brightest $\sim50$ LMC sources). We determine the best fit triangle match using a least squares optimization of all the side-length to perimeter ratios (\hyperref[fig:matching]{Figure 1}). Once we have an initial link between three CGI detector sources and three LMC stars, we compute an initial plate scale and north angle from these sources alone (see \hyperref[subsec:platescale]{Section 3.1}, \hyperref[subsec:northangle]{3.2}). We then translate all the stars in the LMC subfield from their [RA, Dec] sky coordinates to [pixel, pixel] predicted positions on the detector using the \code{astropy}\cite{astropy_2022} World Coordinate System module and our initial plate scale and north angle. Finally, we match the rest of the CGI detector sources by proximity to these predicted positions of sources. 

\section{ASTROMETRIC CALIBRATION}\label{sec:astrometric_calibration}

\subsection{Plate Scale}\label{subsec:platescale}

The plate scale is the conversion metric between distance on the detector [pixels] and angular separation on the sky [mas]. This metric allows us to understand how the detector image scales in relation to the calibration field. For the Roman CGI, we have an initial understanding that the plate scale will be near $21.8$ [mas/pixel] and with the astrometric calibration we will be able to directly measure this value and validate it with real data. We note that the process for calculating plate scale relies heavily on the result of our automated source matching algorithm (\hyperref[sec:finding_and_matching]{Section 2}).

\begin{figure}[ht]
   \begin{center}
   \begin{tabular}{c} 
   \includegraphics[height=8cm]{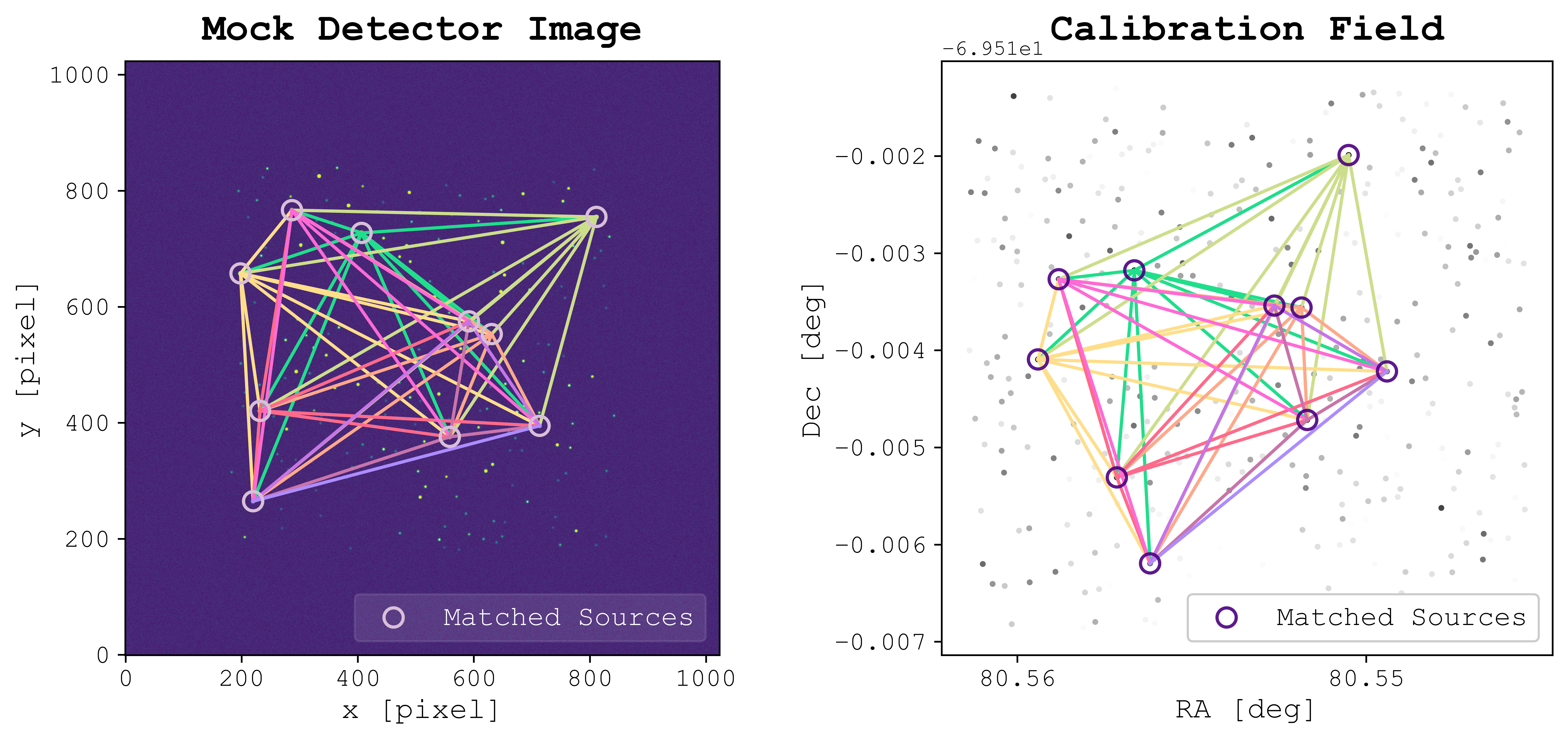}
   \end{tabular}
   \end{center}
   \caption{\label{fig:platescale_northangle} 
\textbf{Left:} A \code{corgidrp} mock detector image showing all the possible combinations (colored lines) of unique star pairs using 10 sources (purple circles). \textbf{Right:} The LMC calibration subfield showing the corresponding combinations of unique star pairs from the matched 10 sources in the left panel.}
\end{figure}

To begin the plate scale calculation, we compute all unique two-star combinations of N stars for which we know both their detector [pixel, pixel] position and corresponding [RA, Dec] coordinates (\hyperref[fig:platescale_northangle]{Figure 2}). We compute the relative offsets between unique star pairs using a custom PSF fitting routine built with \code{scipy}\cite{scipy_2020} subpackages, \code{ndimage} and \code{optimize}, that shifts and fits gaussian psfs to star locations. Additionally, to calculate the true on-sky separation [mas] between each pair of matched stars in the LMC calibration field, we use the \code{coordinates} package in \code{astropy}\cite{astropy_2022}. Finally, we take the ratio of the detector distance [pixels] and on-sky separation [mas] for each of the unique star pairs and report the median of these measurements as the detector plate scale.

\subsection{North Angle}\label{subsec:northangle}

The north angle describes the direction of true North in the CGI detector as it relates to the positive y axis-- allowing us to re-orient the CGI detector image with respect to the observed field. The convention we use to describe the north angle is the counter-clockwise direction from true North\footnote{True North is defined as the positive declination axis in the calibration field} to the positive y axis in the CGI detector image. We note that, based on the CGI optical design, this convention is also used to define East as 90 [deg] counter-clockwise of North. Similarly to the plate scale algorithm, the process for calculating the north angle relies heavily on the result of our automated source matching algorithm (\hyperref[sec:finding_and_matching]{Section 2}).

To begin the north angle calculation, we use all the unique combinations of star pairs for N stars that were determined in \hyperref[subsec:platescale]{Section 3.1} (\hyperref[fig:platescale_northangle]{Figure 2}). For each pair, we calculate the position angle from one star to another, oriented from the detector's positive y axis and sweeping counter clockwise. We then measure the position angle between the corresponding matched star pairs in the LMC calibration field using the \code{astropy}\cite{astropy_2022} \code{coordinates} package. This on-sky position angle is oriented from the positive declination axis and sweeping counter clockwise with the right ascension axis increasing to the left. Finally, we compute the north angle for each star pair as the difference between the on-sky position angles and CGI detector angles and report the median value as the detector north angle.

\subsection{Boresight Calibration}\label{subsec:boresight}

The boresight calibration determines the offset in pointing of the coronagraph and reports the true [RA, Dec] coordinates of the CGI detector center. In coronagraphic observations, the pointing target itself is hidden behind the coronagraph mask. Although we utilize a control loop to place the target exactly behind the mask, we require a good boresight calculation to inform the initial pointing of CGI. Therefore, measuring the boresight allows us to obtain accurate and precise relative astrometry of our science targets.

The boresight calibration relies directly on the plate scale and north angle metrics calculated in \hyperref[subsec:platescale]{Section 3.1} and \hyperref[subsec:northangle]{Section 3.2}. After obtaining the plate scale and north angle of the detector, we use \code{astropy}\cite{astropy_2022} World Coordinate System to translate stars from the LMC calibration field into predicted positions on the CGI detector with respect to the commanded CGI pointing coordinates. In measuring the difference between predicted positions of sources and their found positions on the CGI detector we can determine the pointing offset in both the x and y directions. We can translate this offset from [pixel] units to on-sky [deg] offsets and report the true [RA, Dec] coordinates of the detector center.

\subsection{Optical Distortion Mapping}\label{subsec:distortion}

The optical distortion mapping is a process of measuring and mapping the spatially variant pixel offsets across the CGI detector. Imperfections in instrument optics can lead to the warping of pixels near the edges of the detector image. These aberrations in the CGI detector plate scale and north angle must be measured and removed in order to recover accurate astrometry from science observations. Similarly to the boresight calibration, the optical distortion mapping relies on the plate scale and north angle metrics derived in \hyperref[subsec:platescale]{Section 3.1} and \hyperref[subsec:northangle]{Section 3.2}.

We utilize two-dimensional Legendre polynomials to characterize the detector distortion at each pixel. This methodology is modeled after the NIRC2 distortion solution for the Keck II telescope\cite{Service_2016}. The incorporation of Legendre polynomials is implemented through the \code{numpy}\cite{numpy_2020} Legendre polynomial class and we default to using third order polynomials described by \hyperref[eq:legendre]{Equation 1}. 
\begin{equation}\label{eq:legendre}
    p(x,y) = \sum_{i,j} c_{i,j} * L_i(x) * L_j(y)
\end{equation}
Our distortion mapping returns the multi-degree Legendre polynomial coefficients $c_{i,j}$. For third order polynomials, we return an array of 32 coefficients (16 for each axis) and append the order of Legendre polynomials to the astrometric calibration file produced by \code{corgidrp}.

To compute the optical distortion across the detector, we begin by taking all unique combinations of sources we detect in our source matching algorithm. We measure the separation [pixels] and position angle [deg] (with the same angle convention as in \hyperref[subsec:northangle]{Section 3.2}) between all unique pairs of sources. We also measure the position angle and separation between the unique pairs' on-sky coordinates as a reference, since optical distortion will cause variations in these quantities across the detector. We translate the true on-sky separations from milliarcsecond units to pixels given the derived plate scale (\hyperref[subsec:platescale]{Section 3.1}) and align their orientations to the detector by subtracting the derived north angle (\hyperref[subsec:northangle]{Section 3.2}) from the star pairs' position angle. Next, we un-distort the measured star positions given some initial distortion coefficient parameters. Then, we compute the residual between each of the measured, un-distorted separations and position angles and the on-sky separations and position angles. Finally, we fit for Legendre polynomial coefficients that minimize these residuals using the \code{scipy}\cite{scipy_2020} optimization framework.

\begin{figure}[ht]
   \begin{center}
   \begin{tabular}{c} 
   \includegraphics[height=8cm]{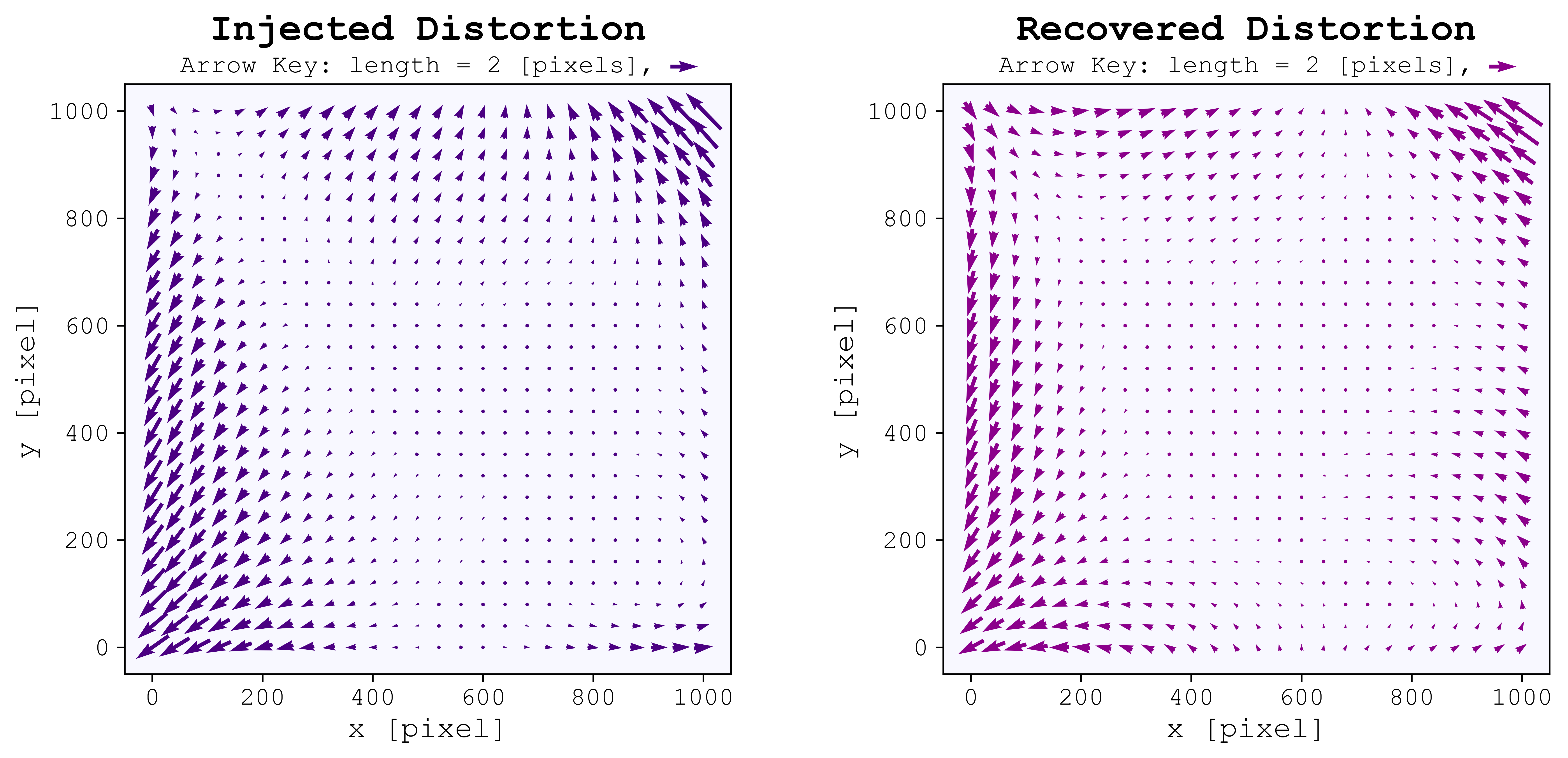}
   \end{tabular}
   \end{center}
   \caption{\label{fig:distortion} 
\textbf{Left:} Map of the injected distortion in \code{corgidrp} mock data with vector arrows of [x, y] offsets for each pixel on the detector and arrow size corresponding to vector strength (key at the top of the figure). \textbf{Right:} Map of the distortion recovered in \code{corgidrp} testing with vector arrows of [x, y] offsets for each pixel on the detector and with arrow size corresponding to vector strength.}
\end{figure}

\section{TESTING}\label{sec:testing}

The testing of our astrometric calibration algorithm is designed to meet the formal, CGI-specific Technology Demonstration Threshold Requirements (TTR5) determined by the program-level requirements of the Roman Space Telescope. The astrometric calibration must meet the following requirements\cite{Zellem_2022}: (1) compute the location of the central pixel of the detector in the ICRF frame to within 30 mas ($3\sigma$) and (2) compute the orientation of the detector’s axes with respect to galactic north to within 0.3 degrees ($3\sigma$). We note that there are no formal requirements for computing the plate scale or distortion map to any threshold, however, we set arbitrary requirements for these metrics within our testing framework. We expect that there will be fewer than 4 [mas] of distortion present in the central 1 arcsecond x 1 arcsecond of the CGI detector and, therefore, require distortion mapping to be accurate to within 4 [mas] in this region. Additionally, we require that plate scale is measured to within 0.5 [mas/pixel] in order to ensure accurate computation of the other astrometric calibration products. 

We test our ability to reach the aforementioned requirements by testing through two avenues: \code{corgidrp}\footnote{https://github.com/roman-corgi/corgidrp} (\hyperref[subsec:corgidrp_testing]{Section 4.1}) and \code{corgisim}\footnote{https://github.com/roman-corgi/corgisim} testing (\hyperref[subsec:corgisim_testing]{Section 4.2}). The first version of testing, with \code{corgidrp}\cite{Wang_2026}, contains the framework to generate simple mock detector images for astrometric calibration testing within the data reduction pipeline itself. The second, most recent, testing pathway uses higher fidelity \code{corgisim}\cite{Millar_Blanchaer_2026} CGI detector simulations as input for testing the astrometric calibration. While the \code{corgidrp} testing is vital to assessing the functionality of the astrometric calibration as a whole, further testing will focus on \code{corgisim} data as it better represents the raw CGI data we expect to receive during the commissioning phase of the mission.

\subsection{Testing with \code{corgidrp} Data }\label{subsec:corgidrp_testing}

The \code{corgidrp} mock detector images model CGI sources as two-dimensional gaussian distributions and include random gaussian noise. In these simulated images, we model the translation of star positions from the LMC calibration field to the CGI detector locations with \code{astropy}\cite{astropy_2022} in the same manner as in \hyperref[sec:finding_and_matching]{Section 2}. Our mock CGI datasets are complex enough to test the functionality of the astrometric calibration, but do not accurately represent raw CGI data. These mock data are shown in the left panel of \hyperref[fig:matching]{Figure 1} and \hyperref[fig:platescale_northangle]{Figure 2}. 

The framework to test individual pipeline functions exists within \code{corgidrp}, but a key aspect of testing is the performance of a calibration algorithm in conjunction with other \code{corgidrp} processing functions. We use \code{corgidrp} end-to-end tests to ensure that our astrometric calibration function can be easily appended to the result of other processing elements in the pipeline. In this way, we are able to process mock astrometric data from their raw (L1) state all the way through the astrometric calibration sequence. The results of \code{corgidrp} end-to-end testing reveal we are able to recover the location of the detector center with an error of $0.723$ [mas] in right ascension and $0.151$ [deg] in declination, as well as recover the orientation of the CGI detector axes with error of $0.018$ [deg]. Aside from the TTR5 requirements, we also measure plate scale with an error of $0.027$ [mas/pixel] and distortion with error of $0.824$ [mas] in the x-axis and $0.790$ [mas] in the y-axis in the central 1 arcsecond x 1 arcsecond region of the detector.

\subsection{Testing with \code{corgisim} Data }\label{subsec:corgisim_testing}

The most recent method of astrometric calibration testing relies on \code{corgisim}\cite{Millar_Blanchaer_2026}, a python-based simulation suite for Roman CGI. The use of \code{corgisim} allows us to test our astrometric calibration algorithm on simulations that better represent raw Roman coronagraphic observations by incorporating more realistic PSFs, image vignetting, cosmic ray effects, and detector noise patterns. Efforts to run the \code{corgisim} data through full astrometric calibration end-to-end testing within \code{corgidrp} are still in progress, however, we are able to run initial individual processing and testing sequences on a \code{corgisim} dataset. 

\begin{figure}[ht]
   \begin{center}
   \begin{tabular}{c} 
   \includegraphics[height=8cm]{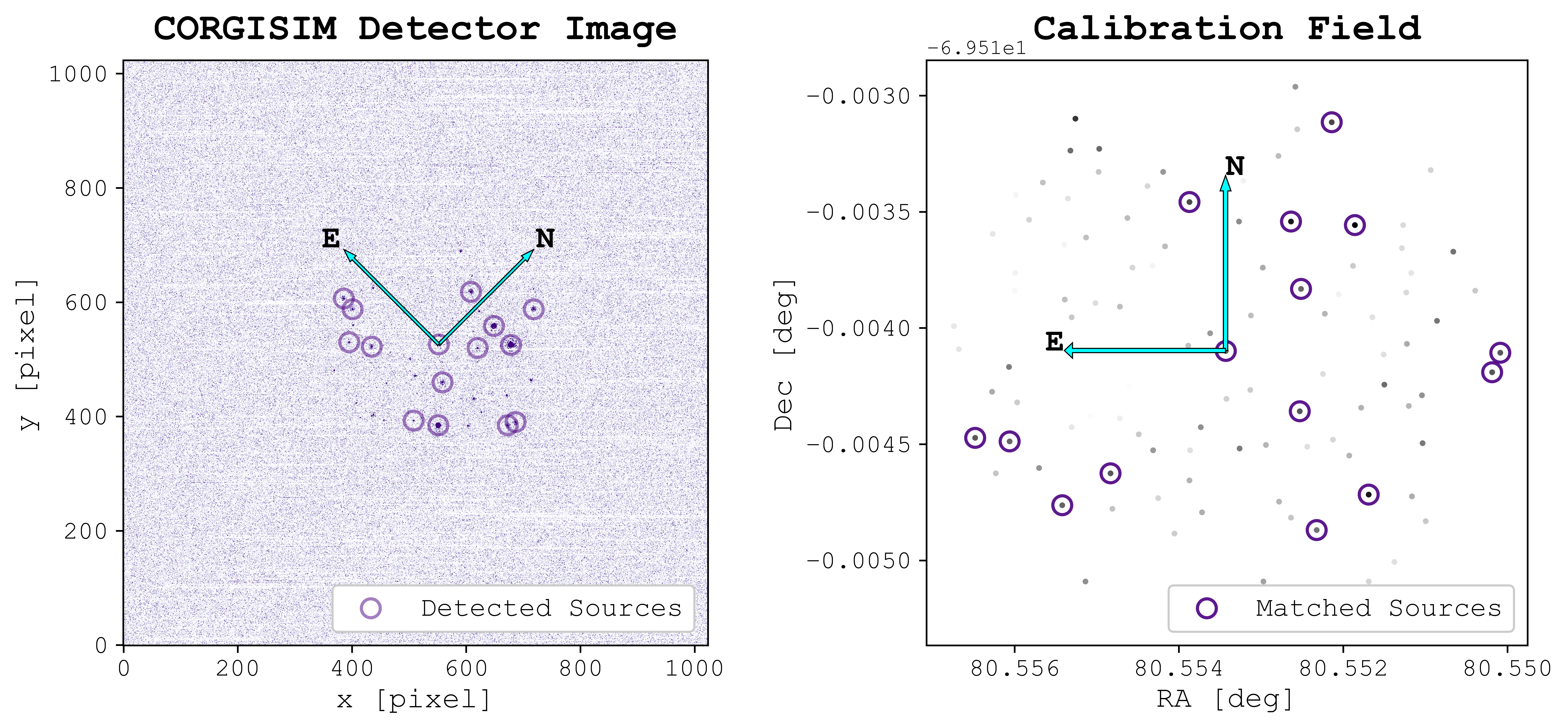}
   \end{tabular}
   \end{center}
   \caption{\label{fig:corgisim_testing} 
\textbf{Left:} The semi-processed \code{corgisim} data with detected sources circled in purple and the recovered north angle orientation shown with cyan arrows. \textbf{Right:} The LMC calibration field with matched sources circled in purple and the true North and East orientation shown with cyan arrows.}
\end{figure}

We utilize individual processing steps within the \code{corgidrp} L1 to L2 reduction procedure to prepare \code{corgisim} data for astrometric calibration without relying on the success of other processing steps. In particular, we measured and subtracted the median bias of the pre-detector scan and cropped the image to the science region. Additionally, we median combined \code{corgisim} frames to mitigate cosmic ray appearance and utilized the cosmic ray detection algorithm create a mask of the rays. We then multiplied the mask and detector image together before subtracting this product from the image itself-- sufficiently eliminating remaining cosmic ray traces. After performing these individual processing steps, we were able to run the astrometric calibration on the \code{corgisim} data (\hyperref[fig:corgisim_testing]{Figure 4}). We recover the plate scale with $0.242$ [mas/pixel] error and north angle with $0.004$ [deg] error. 

\section{CONCLUSION}\label{sec:conclusion}

We present the Roman \code{corgidrp} astrometric calibration algorithm in all of its parts: automated source matching, plate scale calculation, north angle calculation, boresight calibration, and optical distortion mapping. We have shown that the algortihm performs to standards of the official Roman Technology Demonstration Threshold Requirements with \code{corgidrp} mock astrometric data simulations. We will continue efforts to test our astrometric calibration on higher fidelity \code{corgisim} simulations to ensure the algorithm's robustness. The next phase of implementing the astrometric calibration will be with real data from the Roman Space Telescope in the coming months.

\acknowledgments       
 
This research was carried out in part at the Jet Propulsion Laboratory, California Institute of Technology, under a contract with the National Aeronautics and Space Administration (80NM0018D0004). We would like to thank and acknowledge contributions from Sergi Hildebrandt, formerly at NASA Jet Propulsion Laboratory. This material is based upon work supported by the National Aeronautics and Space Administration under award No.80NSSC24K0087.

% References
\bibliography{report} % bibliography data in report.bib
\bibliographystyle{spiebib} % makes bibtex use spiebib.bst

\end{document}